# Investigating of electrons bunching in a Penning trap and accelerating process for $CO_2$ gas mixture active medium

Tian Xiu-fang(田秀芳)[1]　　Wu Cong-feng(吴丛凤)[1]*　　Jia Qi-ka(贾启卡)[1]

[1](National Synchrotron Radiation Laboratory, University of Science and Technology of China, Hefei 230029, China)

**Abstract**: In the presence of an active medium incorporated in a Penning trap, the moving electrons can become bunched, as they get enough energy, they escape the trap forming an optical injector. These bunched electrons can enter next PASER section filled with the same active medium to be accelerated. In this paper, electron dynamics in the presence of gas mixture active medium incorporated in a penning trap is analyzed by developing an idealized 1D model. We further evaluate the energy exchange occurring as the train of electrons traversing the next PASER section. The results show that the oscillating electrons can be bunched at the resonant frequency of the active medium. The influence of the trapped time and the population inversion are analyzed, which shows that the longer the electrons are trapped, the more energy from the medium the accelerated electrons get, and with the increase of the population inversion, the decelerated electrons virtually unchanged but the accelerated electrons more than double their peak energy values. What is more, the simulation results show that the gas active medium need lower population inversion to bunch the electrons compared to the solid active medium. So the experimental condition is easy to be achieved.

## 1 Introduction

In the particle acceleration by stimulated emission of radiation (PASER) [1-3], the energy being stored in molecules or atoms is directly used to accelerate electrons. PASER doesn't need for phase-matching or compensating for phase slippage. And it doesn't require high power laser or beam driver; further more an electron gun is also not needed. And recently, the calculated results show that the wake generated by the trigger bunch of electrons in the active medium can get 1GV/m [4-6].

In the traditional accelerator structure, phase-matching is necessary which ensures the particle only experiences the field pointing in the same direction as the particle motion, thereby, allowing it to gain net energy. But in PASER, the electrons interact with the active medium via a virtual photon. This photon is emitted by the excited molecule, therefore, when the electron absorbs this photon, the electron gains energy equal to the energy emitted when the electron of the excited molecule returns from upper to lower energy state. The electric field of this virtual photon is not critical for this absorption process to occur. Hence, phase-matching is not needed.

In the PASER, the electron passing nearby the excited molecule stimulates the molecule to emit a photon, which is absorbed directly by the electron. If the electrons are bunched together with a spacing equal to the wavelength of the emitted photons, i.e., at the resonance of the excited state, then there is a coherent effect that further enhances the energy exchange process. This is why it is preferable to have the electrons bunched at the resonance wavelength during the PASER process.

In the proof-of-principle PASER experiment [3]

[1]Supported by National Natural Science Foundation of China (10675116) and Major State Basic Research Development Programme of China (2011CB808301)
*通讯作者：吴丛凤，**E-mail**: cfwu@ustc.edu.cn
作者简介：田秀芳，女，（1983-），博士生，主要研究方向为新加速原理和新加速结构。
**E-mail**: txiufang@mail.ustc.edu.cn

which gave the first experimental result about PASER. The bunch equipment consists of existing accelerator, wiggler, and high power laser, which are very complex and expensive. In order to replace these three components, Dr. Levi Schachter suggested a novel paradigm [7,8] which relies on the possibility that in the presence of solid active medium (Nd:YAG), the non-relativistic electrons oscillating in a Penning trap may get bunched at the resonance frequency of the active medium. During multiple round trips in the trap, the bunched electrons get enough kinetic energy, and can escape the trap forming a low energy optical injector. Thus, electron gun is not needed.

When the bunched electrons escape the trap, they can enter the next PASER section to get higher energies.

We have made some theory analysis and simulations about PASER, some results are given in the paper [9], and in this paper, electron dynamics in the presence of $CO_2$ gas mixture active medium incorporated in a Penning trap is analyzed, fig.1 shows the schematics of a Penning trap. Compared to the solid active medium, the $CO_2$ gas mixture active medium is less costly and easy to get. In the following, based on the 1D solid active medium model [8], we develop an idealized 1D model in gas mixture active medium to analyze the bunching process in the penning trap and further make calculations by MATLAB, the influence of the oscillating time in the trap and the population inversion are investigated. Furthermore, calculated simulation of kinetic energy gain of the bunched electrons in the next PASER section is studied.

## 2 Bunching process of electrons in the presence of $CO_2$ gas mixture active medium in a penning trap

Based on the 1D solid active medium model [8], consider the difference of dielectric function $\varepsilon$ and the "plasma" frequency $\omega_p$ between the solid and gas active mediums, electron dynamics in the presence of $CO_2$ gas mixture active medium incorporated in a Penning trap is investigated by deducing an idealized 1D model and simulations.

The dielectric function of gas active medium is expressed by

$$\varepsilon(\omega > 0) \equiv 1 + \frac{\omega_p^2}{\omega_0^2 - \omega^2 + 2j\omega/T_2} \quad (1)$$

Which satisfy $\varepsilon(\omega < 0) = \varepsilon^*(\omega > 0)$. It is assumed that the medium has a single resonance frequency chosen to correspond to the macro-bunch modulation $\omega_0 = 2\pi c/\lambda_0$, $\omega_p$ is the "plasma" frequency, $\omega_p^2 \equiv e^2 \Delta n/m\varepsilon_0$, with m being the rest mass of the electron and $\Delta n$ representing the population density of the resonant atoms. For an excited medium, when the population density is inverted (n<0), the plasma frequency is negative ($\omega_p^2 < 0$); $T_2$ is the relaxation time.

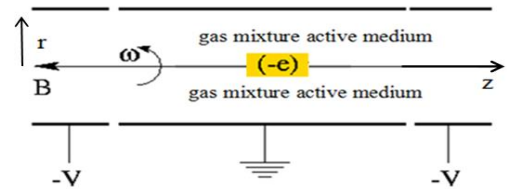

Fig.1 Schematics of a Penning trap. The trap uses coils to generate a uniform axial magnetic field to provide radial confinement and applied end potentials provide axial confinement.

Assuming that, in the penning trap, there are $N_{mp}$ macro-particles, and each macro-particle contains $N_{el}$ electrons. If the vth macro-particle's trajectory is represented by ($r_v(t), z_v(t)$), then the longitudinal current density can be expressed by:

$$J_z(r,z,t) = -eN_{el} \sum_v \dot{z}_v(t) \frac{1}{2\pi r} \delta[r - r_v(t)]\delta[z - z_v(t)] \quad (2)$$

The magnetic potential associated with the above current density in the boundless case is given by

$$A_z(r, k, \omega) = \mu_0 \int dr' r' G(\Lambda r, \Lambda r') J_z(r', k, \omega) \quad (3)$$

In which $J_z(r', k, \omega)$ is the spatial and temporal Fourier transform of the longitudinal current density in equation (2). And $\Lambda^2 = k^2 - \varepsilon(\omega)\omega^2/c^2$

$$G(\Lambda r, \Lambda r') = \begin{cases} I_0(\Lambda r) K_0(\Lambda r'), r < r' \\ K_0(\Lambda r) I_0(\Lambda r'), r > r' \end{cases} \quad (4)$$

So we can get the longitudinal electric field generated by the ensemble electrons in the trap as follows:

$$E_z(r, z, t) = \frac{\mu_0 e N_{el} N_{mp}}{(2\pi)^3} \int_{-\infty}^{\infty} d\omega \exp(j\omega t)$$
$$\times \int_{-\infty}^{\infty} dk \exp(-jkz) \frac{c^2 \Lambda^2}{j\omega \varepsilon(\omega)}$$
$$\times \int_{-\infty}^{\infty} dt' \exp(-j\omega t')$$
$$\times \langle G[\Lambda r, \Lambda r_\nu(t')] \dot{z}_\nu(t') \exp[jkz_\nu(t')] \rangle_\nu \quad (5)$$

Therefore, the force on a single macro-particle (the $\nu$th macro-particle) is expressed by

$$F_\nu(t) = -eN_{el} E_z[r_\nu(t), z_\nu(t), t]$$
$$= -\frac{\mu_0 e^2 N_{el}^2 N_{mp}}{(2\pi)^3} \int_{-\infty}^{\infty} d\omega \exp(j\omega t)$$
$$\times \int_{-\infty}^{\infty} dk \exp[-jkz_\nu(t)] \frac{c^2 \Lambda^2}{j\omega \varepsilon(\omega)} \int_{-\infty}^{\infty} dt' \exp(-j\omega t')$$
$$\times \langle \langle G[\Lambda r_\nu(t), \Lambda z_\mu(t')] \rangle \dot{z}_\mu(t') \exp[jkz_\mu(t')] \rangle_\mu \quad (6)$$

And the total energy exchange is given by

$$W_{ex} = N_{mp} \int_{-\infty}^{\infty} dt \langle \dot{z}_\nu(t) F_\nu(t) \rangle_\nu$$
$$= -\frac{\mu_0 e^2 N_{el}^2 N_{mp}^2}{(2\pi)^3} \int_{-\infty}^{\infty} dt \int_{-\infty}^{\infty} d\omega \exp(j\omega t)$$
$$\times \int_{-\infty}^{\infty} dk \frac{c^2 \Lambda^2}{j\omega \varepsilon(\omega)} \int_{-\infty}^{\infty} dt' \exp(-j\omega t')$$
$$\times \left\langle \begin{array}{c} \dot{z}_\nu(t') \exp[-jkz_\nu(t)] G[\Lambda r_\nu(t), \Lambda r_\mu(t')] \\ \times \dot{z}_\mu(t') \exp[jkz_\mu(t')] \end{array} \right\rangle_{\nu,\mu} \quad (7)$$

Before proceeding, we make the following assumptions: (1) the transverse distribution is independent of the longitudinal distribution, and it's contribution to the energy exchange is negligible, (2) in the radial direction, the electrons are uniformly distributed in the range $0<r<R_b$, so the transverse filling factor $\langle G[\Lambda \rho_\nu, \Lambda \rho_\mu] \rangle_{\nu,\mu}$ can be replaced by transverse form factor $F_\perp(\Lambda R_b) = 2[1 - 2K_1(\Lambda R_b)I_1(\Lambda R_b)]/(\Lambda R_b)^2$, for $CO_2$ gas mixture active medium, the transverse wavelength is much larger than the radius of the electrons' ensemble, so $\Lambda R_b$ is very small, and we can approximate $[1 - 2K_1(\Lambda R_b)I_1(\Lambda R_b)] \simeq (\pi x/2)^2$.

In the penning trap, the angular frequency $\Omega$ is determined by $\Omega = (2c/L)\sqrt{2eV_0/mc^2}$, L is the length of the penning trap, so ignoring the damping during one period of the oscillation, the trajectory of the $\nu$th macro-particle can assumed to be given by

$$z_\nu(t) = \frac{L}{2}\{1 + \cos[\Omega(t - t_\nu)]\} \quad (8)$$

So the velocity can be expressed by

$$\dot{z}_\nu(t) = -\frac{L}{2}\Omega \sin[\Omega(t - t_\nu)] \quad (9)$$

Therefore, after a series of derivation, the total energy exchange is as follows:

$$W_{ex} = \frac{-e^2 N_{el}^2 N_{mp}^2 (2\pi)^3}{4\pi\varepsilon_0 \Omega^2 (2L)} \frac{e^2 \Delta n}{m\varepsilon_0} \langle \cos[\omega_0(t_\nu - t_\mu)]$$
$$\times \exp\left[-\frac{1}{T_2}(t_\nu - t_\mu)\right] h(t_\nu - t_\mu) \rangle_{\nu,\mu} \quad (10)$$

So the force on the $\nu$th macro-particle contributing directly to the energy exchange is

$$F_\nu(t) = \frac{2e^2 N_{el} N_{mp}(2\pi)^2}{4\pi\varepsilon_0 \Omega^2 L^2} \frac{e^2 \Delta n}{m\varepsilon_0} \langle \cos[\omega_0(t_\nu - t_\mu)]$$
$$\times \exp\left[-\frac{1}{T_2}(t_\nu - t_\mu)\right] h(t_\nu - t_\mu) \rangle_\mu \sin[\Omega(t - t_\nu)] \quad (11)$$

And the amplitude of the force is

$$f_\nu(t) = \frac{2e^2 N_{el} N_{mp}(2\pi)^2}{4\pi\varepsilon_0 \Omega^2 L^2} \frac{e^2 \Delta n}{m\varepsilon_0} \langle \cos[\omega_0(t_\nu - t_\mu)]$$
$$\times \exp\left[-\frac{1}{T_2}(t_\nu - t_\mu)\right] h(t_\nu - t_\mu) \rangle_\mu \quad (12)$$

Assuming the effective impact of the trap on the particles is represented by an ideal harmonic oscillator, then its force is $\Omega^2 z$, and the equation of motion of the $\nu$th macro-particle is

$$F_\nu(t) + F_{scatt} + \Omega^2 z_\nu = N_{el} m \ddot{z}_\nu \quad (13)$$

In which $F_{scatt}$ is the effect of the elastic collisions of the electrons with gas.

Explicitly, the equation of motion can be expressed by

$$\frac{d^2 z_\nu}{dt^2} + \left(\frac{2}{\tau_\nu^{(csk)}} + \frac{2}{\tau_{scatt}}\right)\frac{dz_\nu}{dt} + \Omega^2 z_\nu = 0 \quad (14)$$

In which the decay parameter $\tau_\nu^{(csk)}$ is represent by

$$\frac{2}{\tau_\nu^{(csk)}} = -\frac{2\pi^2 \Delta n e^2 r_e N_{mp} N_{el}}{\varepsilon_0 mc(2eV_0/mc^2)^{3/2}} \times \langle cos[\omega_0(t_\nu - t_\mu)] exp\left[-\frac{1}{T_2}(t_\nu - t_\mu)\right] h(t_\nu - t_\mu)\rangle_\mu \quad (15)$$

Assuming that, during one round trip in the trap, the decay parameter doesn't change significantly, solving the equation of motion, we can get that after one round trip, the $\nu$th macro-particle's phase space can be given by

$$z_\nu(T) = z_\nu(0)exp(-T/\tau_\nu) \quad (16)$$
$$\dot{z}_\nu(T) = \dot{z}_\nu(0)exp(-T/\tau_\nu) \quad (17)$$

In which $(z_\nu(0), \dot{z}_\nu(0))$ is the initial phase space. For $CO_2$ gas mixture active medium, $\lambda_0 = 10.6\mu m$, assuming the population density is $\Delta n \sim 10^{11} m^{-3}$, the relaxation time $T_2 = 2ms$, assume the trap is of length L=0.01m, and the voltage $V_0$ on the central anode is 400V, $N_{mp} = 500$, $N_{el} = 10^6$, and the decay time $\tau_{scatt}$ is assumed to be 0.08ms. Based on the parameters and the equations above, we make numerical simulations to trace a fraction of the ensembles that populates one optical period and further assume that there is no significant difference from one period to another.

Fig.2 shows the amplitude of the force on the $\nu$th macro-particle in one optical period. We can see that during one optical period some electrons are accelerated while others are decelerated.

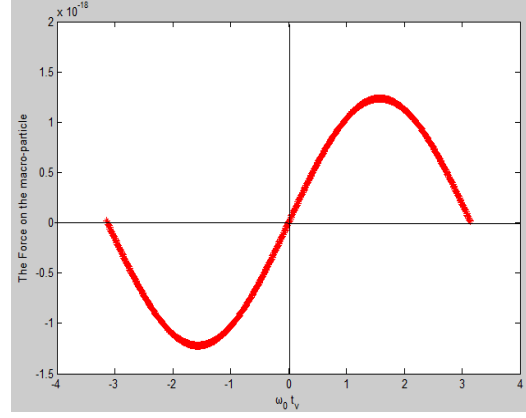

Fig.2 The amplitude of the force on the $\nu$th macro-particle in one optical period.

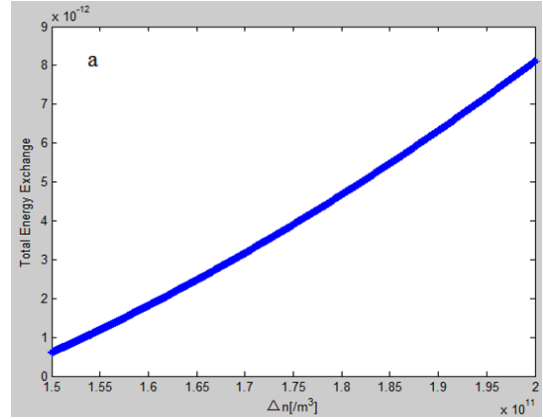

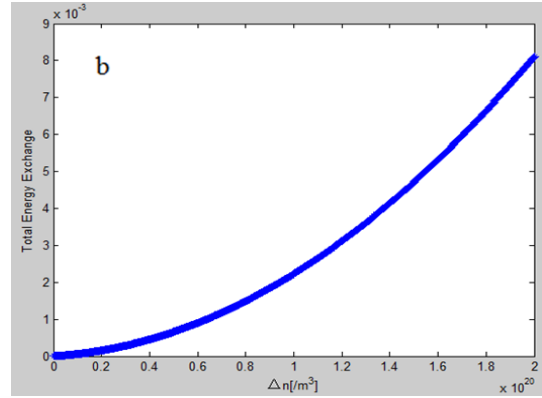

Fig.3 The total energy exchange versus the population inversion

The variation of the total energy exchange versus the population inversion is illustrated in fig.3, a shows the lower population inversion region while b illustrate the higher region. The total energy exchange is not linearly proportional to the population inversion but is increased with the increasing of the population inversion.

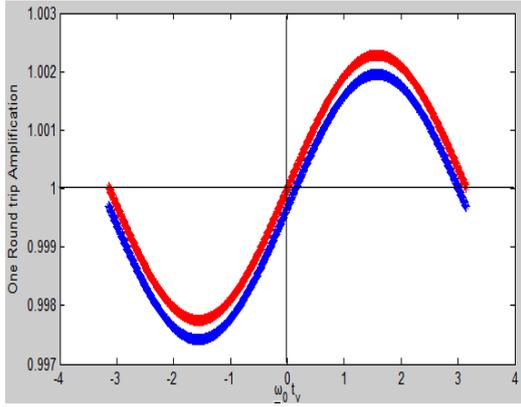

Fig.4 Amplification factor $\exp(-T/\tau_\nu)$ with (blue line) and without (red line) scattering included.

We considered the one round trip amplification factor $\exp(-T/\tau_\nu)$ with and without regular scattering in fig.4, which show that, in the absence of normal scattering, a significant fraction of electrons can absorb energy from the medium, as the scattering effect increases, this fraction diminishes. In order to envision the impact of this effect, we show in fig.5 the relative changes in the total energy of the electrons with and without normal scattering after 500 round trips. We observe that in the absence of the normal scattering, a significant fraction of electrons can absorb energy from the active medium, primarily those around phase $\pi/2$.

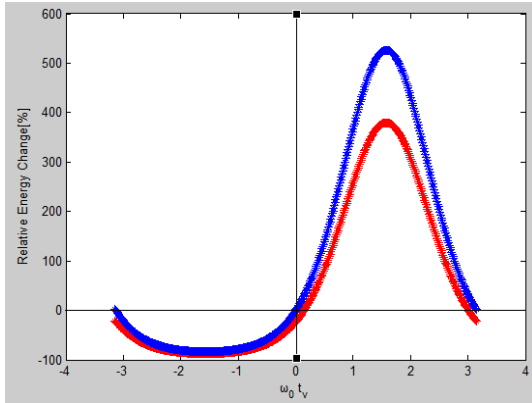

Fig.5 The relative changes in the total energy of the electrons with (the red line) and without (the blue line) normal scattering after 500 round trips.

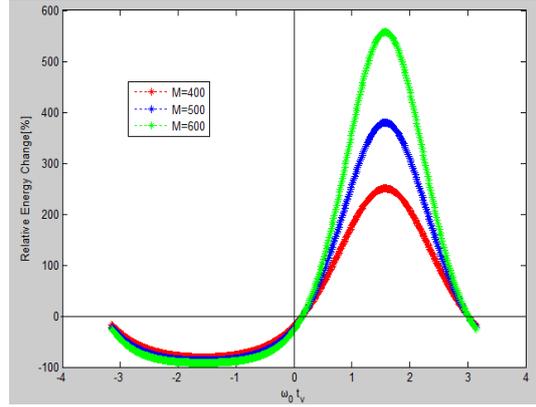

Fig.6 The relative energy changes for 400,500,600 roundtrips with normal scattering. The population inversion is $\Delta n \sim 2 \times 10^{11} \mathrm{m}^{-3}$

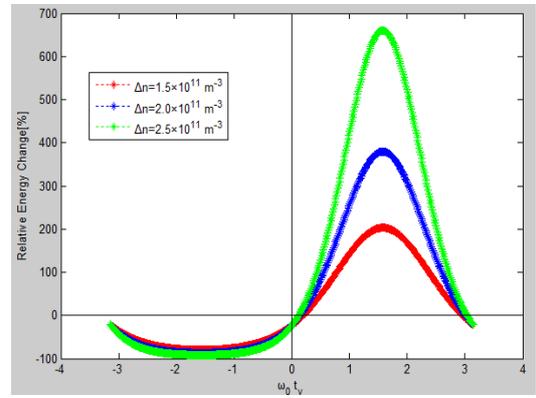

Fig.7 The relative energy changes for 500 roundtrips with different population inversion.

When the population inversion is $\Delta n \sim 2 \times 10^{11} \mathrm{m}^{-3}$, the relative energy changes after 400,500,600 roundtrips with normal scattering are showed in fig.6. This figure illustrates that the longer the electrons are trapped, the more energy from the medium the accelerated electrons get, but at the same time, the energy of the decelerated electrons various slowly.

Fig.7 indicates the case that with different population inversion, the relative energy changes after 500 roundtrips, which shows that with the increase of the population inversion, the decelerated electrons virtually unchanged but the accelerated electrons more than double their peak energy values.

It's important to note that the above theory simulation results show that the gas active medium need lower population inversion to bunch the electrons compared to the solid active medium, so the experimental condition is easy to be achieved.

# 3 Accelerating process of kinetic energy gain of bunched electrons in $CO_2$ gas mixture active medium

When the oscillating electrons get bunched and gain enough energy from the active medium, they escape from the trap. The electrons can transverse into the next acceleration unit which is filled with the same gas mixture active medium to be accelerated in order to reach high energies. The schematic is illustrated in fig.8.

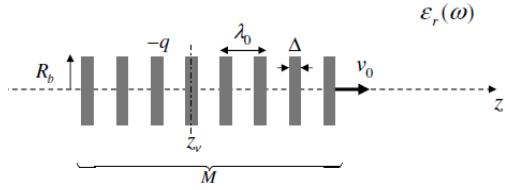

Fig.8 A macro-bunch consisting of a train of *M* micro-bunches traversing a medium characterized by its dielectric coefficient $\varepsilon_r$ (ω). Each micro-bunch carries a charge $-q$.

The energy exchange between the bunched electrons and the active medium can be expressed by [9]

$$W = \frac{Q^2 d}{4\pi\varepsilon_0 \lambda_0^2 \beta^2} \frac{\Omega_p^2}{\Omega_R} F_\parallel \left(\frac{\Omega_R}{2\beta}, \bar{\Delta}, M\right)$$
$$\times \text{Re}\left[\Omega_+ F_\perp \left(\frac{\Omega_+ \bar{R}_b}{\beta}\right)\right] \quad (18)$$

Wherein $\bar{R}_b = R_b/\lambda_0, \bar{\Delta} = \Delta/\lambda_0, \Omega = \omega\lambda_0/c, \varphi = \Omega/2\beta, u = \Omega\bar{R}_b\sqrt{1-\beta^2\varepsilon_r(\Omega)}/\beta$, are the normalized
quantities, $F_\parallel(\varphi, \bar{\Delta}, M) = \text{sinc}^2(\varphi\bar{\Delta})\text{sinc}^2(\varphi M)/\text{sinc}^2(\varphi)$ ,
$\Omega = \Omega_\pm = j\alpha \pm \sqrt{(2\pi)^2 + \Omega_p^2 - \alpha^2} = j\alpha \pm \Omega_R$ are the poles of the dielectric function, $\alpha = \lambda_0/cT_2$, $\Omega_p^2 = \omega_p^2 \lambda_0^2/c^2$.

It is obvious that, the total energy exchange is proportional to $\Omega_p^2$. Because of the population in the medium is inverted, $\Omega_p^2$ is negative, so the total energy exchange W is negative, therefore, the total kinetic energy gain of the macro-bunch ($\Delta E_k = -W$) is positive, which indicates that energy is transferred from the active medium to the macro-bunch. So the relative kinetic energy change of the macro-bunch reads

$$\overline{\Delta E_k} = \frac{\Delta E_k}{N_{mb} mc^2(\gamma - 1)}$$
$$= \frac{4N_{mb} d(\pi r_e^2)}{\beta^2(\gamma - 1)\Omega_R} \frac{w_{act}}{\hbar\omega_0}$$
$$\times F_\parallel \left(\frac{\Omega_R}{2\beta}, \bar{\Delta}, M\right) \text{Re}\left[\Omega_+ F_\perp \left(\frac{\Omega_+ \overline{R_b}}{\beta}\right)\right] \quad (19)$$

In which $w_{act} = -\Delta n \hbar\omega_0$ is the density of the energy stored in the active medium.

In the acceleration unit, the value of the population inversion should be as large as possible. In the following simulation, the population inversion is assumed to be of the order $\Delta n \sim 10^{23} m^{-3}$. $N_{mb}$ is the total number of electrons in the macro-bunch.

From the expression of the relative kinetic energy change we know that the energy change is linearly proportion to the interaction length d, in the simulation process, we let d=1m.

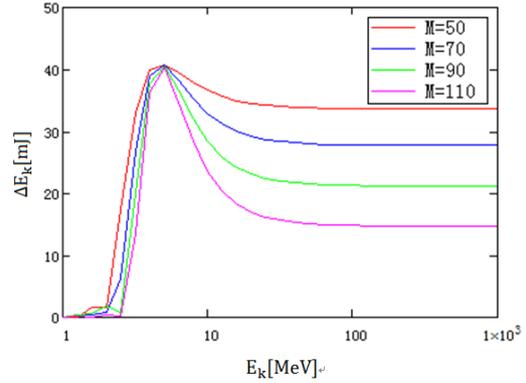

Fig.9 The macro-bunch kinetic energy increase versus the average kinetic energy of the electrons with the number of the micro-bunches as a parameter and $w_{act}$=2000J/m$^3$.

From fig.9 we can see that the energy exchange between the bunched electrons and the active medium is influenced by the initial kinetic energy of the bunched electrons, with the increase of the energy of the bunched electrons, the energy exchange have a peak value, and when the initial kinetic energy of the bunched electrons is large enough, the energy exchange is independent of γ.

## 4 Conclusion and discussion

From the numerical simulations above, we can see that it is possible to use an excited gas mixture inside the Penning trap to make the oscillating electrons to be bunched. When the bunched electrons escape from the trap, they can transverse into the next acceleration unit which is filled with the same gas mixture active medium to be accelerated in order to reach higher energies.

The simulation results show that the longer the electrons are trapped, the more energy from the medium the accelerated electrons get, and with the increase of the population inversion, the decelerated electrons virtually unchanged but the accelerated electrons more than double their peak energy values. A significant fraction of electrons can absorb energy from the medium, as the scattering effect increases, this fraction diminishes. The total energy exchange is linearly proportion to the interaction length and is influenced by the initial bunch kinetic when the bunch energy is low and independent of γ for relativistic energies.

Because PASER is truly a direct photon-to-electron acceleration process, it does not rely on intermediary medium, such as plasma, to accelerate the electron. This is one of the ways it is fundamentally different from plasma accelerators. The advantage is PASER is a simpler acceleration technique that is potentially more compact. PASER does not require a terawatt laser or high-energy electron beam as the driver for generating the plasma wave. Specially, PASER doesn't need phase matching or compensating for phase slippage, and the total energy exchange is linearly proportion to the interaction length. So we may extend the interaction length by bending the electrons' trajectory and re-circulating them in the same cell in order to get high energies.

## 5 Acknowledgments

Thanks Dr. Wayne Kimura for his useful conversations and assistance which made great help for me. Thanks Dr. Levi Schachter for his help.